\documentclass{article}
\usepackage[utf8]{inputenc}
\usepackage{listings}
\usepackage{algorithm}
\usepackage[noend]{algpseudocode}

\algrenewcommand\algorithmicindent{1.2em}
\usepackage{color, colortbl}
\definecolor{Gray}{gray}{0.9}
\newcommand{\bettersim}{{\raise.17ex\hbox{$\scriptstyle\sim$}}}
\usepackage{multirow}
\usepackage{enumitem}
\usepackage{cite}
\usepackage{graphicx}
\usepackage{url}
\usepackage{float}
\usepackage{caption}

\definecolor{dkgreen}{rgb}{0,0.6,0}
\definecolor{mauve}{rgb}{0.58,0,0.82}

\title{Assisted Specification of Code Using Search}
\author{
Steven P. Reiss\\
  Brown University\\
  Providence, RI, USA\\
  \texttt{spr@cs.brown.edu}
}
\date{}

\begin{document}

\maketitle

\begin{abstract}

We describe an intelligent assistant based on mining existing software repositories to
help the developer interactively create checkable specifications of code. 
To be most useful we apply this at the subsystem level, that is chunks of code of 1000-10000 lines that 
can be standalone or integrated into an existing application to provide additional functionality 
or capabilities.  The resultant specifications include both a syntactic description of what should
be written and a semantic specification of what it should do, initially in the form of test cases.
The generated specification is designed to be used for automatic code generation using various
technologies that have been proposed including machine learning, code search, and program synthesis.
Our research goal is to enable these technologies to be used effectively for creating subsystems
without requiring the developer to write detailed specifications from scratch.

\end{abstract}


\maketitle

\section{Motivation}
\label{sec:intro}

Our initial goal is to provide a tool to assist the developer in creating complex, informal, yet checkable 
specifications for software.  Checkable specifications define a software component so that there is a means of 
determining if the software does what is expected.  For many purposes, the expected behavior is good enough.  If
the software is existing or generated by a trusted tool and behaves properly on standard examples,
it is a good starting point.  For example, an agile development sprint can start with a set of interfaces to
implement along with test cases to specify the basic functionality of those interfaces.

We are interested in assisting the developer in building checkable specifications for non-trivial 
software components of between about 1 and 10 KLOC of code we call \emph{subsystems}.  
Smaller components are generally single methods which are fairly easy to specify.  Larger components
bring additional problems to be addressed later.

Such specifications, beyond common tasks such as agile development, will become increasingly important.
Codex has demonstrated that machine learning techniques can be used to 
generate code \cite{codexpaper1}; advanced code
search results demonstrating code reuse from repositories can also be used \cite{scsbcs}; program synthesis is 
becoming increasingly powerful \cite{rbsynguria}.  The initial definition of what should be built is
often informal and ambiguous, for example one might ask for an embedded HTML server or a contact manager.
A variety of quite different results can be generated.  
A practical system using these approaches would 
need to better understand what the developer wanted to better direct the generation,
to choose appropriately from multiple results, and to do validation on the results.  
Moreover, specifications that are close or related
to existing code will typically yield better results with both machine learning and code search since
the prior inputs are likely to cover something close.

With current technology a developer needs to create these specifications manually, for example defining the set
of classes and interfaces along with their methods.  Moreover, they need to provide semantics for these,
defining what the methods do in a checkable manner, typically by providing test cases, but also possibly
with contracts, UML state or event diagrams, or even formal mathematics.  This can be difficult and
creative work that is important to get right. 

For many problems, other developers have had to create a similar specification as part of another system.  
Going through our own systems, we estimated conservatively that 25\% of the code could be reused in
or from other systems.  
Our approach involves mining software repositories to find similar specifications
as a starting point, let the developer choose among and edit those specifications 
to match their particular needs, then use
the edited specification as the basis for mining the software repositories to find 
or generate test cases that can
become part of a checkable result.  
This approach in encompassed in the ASCUS (Assisted Specification of Code Using Search) framework for Java.

ASCUS makes use of several innovative techniques in developing specifications.  
It provides a means for searching for and isolating
compilable subsystems from existing code in repositories.  It offers heuristics that seem to be effective
in providing simplified abstractions of subsystems.  It demonstrates that the type of adaptation that
developers do to make external code meet their requirements can be automated using transformations.  It
shows how test cases can be generated effectively using code search for a particular subsystem specification.

ASCUS, combined with automatic code generation technologies, has the potential to transform software development
by easing much of the burden from the developer, by making code reuse easier, and by building new software
using lessons learned form existing software.  It can provide a practical framework for reusing non-trivial
software components by automatically adapting them to the particular needs of a new application.  It can
form the basis for practical automatic programming.  Our eventual goal is to use ASCUS as the basis
for such code generation.

\section{Related Work}

Our overall efforts build on top of and are related to a variety of different efforts.  
The work that is closest 
to generating whole systems using a step-by-step approach involves 
Model-Driven Development (MDD) \cite{frmddcs,Pastor08}.  In MDD the user writes the system 
using a variety of design notations provided by UML (class diagrams, sequence diagrams, state charts), 
and the system generates code from those specifications.  While this approach is interesting and has a 
growing community of practitioners and researchers, it does not accomplish our objectives.  It does 
not address full applications or the addition of subsystems to an existing application.  With MDD 
the programmer is still writing the code, albeit at a slightly higher level and the specification is 
procedural. Finally, the resultant systems do not make use of the large body of existing 
code and implementations and their embedded experience and knowledge. 

This work is also closely
related to reverse engineering and model-driven reverse engineering \cite{7997723} 
where the purpose is to understand existing systems.  This is
often done for larger systems with graphic models \cite{Muller92,BRUNELIERE20141012}. 
Work has also been done on simplifying specifications for program synthesis \cite{KRASANAKIS}.

There has been extensive work on code search, generally using information 
retrieval and aiming for either code fragments or methods.  While today’s 
repository code search engines are keyword-based, many other approaches have been tried.  
Sourcerer \cite{Bajracharya06} incorporates program structure and semantics in the search base, 
SNIFF incorporates knowledge about libraries and APIs \cite{Chatterjee09}, 
and newer efforts use semantics \cite{Kashyap17,Su16}, test cases \cite{Nurolahzade13}, 
or machine learning \cite{Niu17,clkscwdlmcs}.    
CodeGenie \cite{Lemos11} lets the user define test cases as part of the development 
process in Eclipse and then uses the method names and signatures from the test case 
to build a search query for an internal search engine.  
Wang uses topic-enhanced dependence graphs \cite{Wang11}.
Code recommendation based on 
source code has also studied at the fragment \cite{Holmes08,zjkkghbing,5235134} and 
component level \cite{2903492}. 

Our work on building abstractions of the retrieved code is loosely related to work on 
automated domain model extraction which attempts to extract a domain model from natural 
language requirements \cite{Arora2019AnAL}.  It is more closely related to work on displaying student 
code examples \cite{Head17,Head18}, simple API usage \cite{Glassman18}, or variation in 
small programs \cite{Glassman15}, all
abstracted over a large set of programs. Typical abstraction techniques, such as UML, 
provide another alternative, but are geared more toward the design of a single system 
rather than showing a set of abstractions.  Using UML to describe software product lines \cite{Gomaa04}, takes 
a step towards abstraction, but is limited in scope and flexibility.  Finally, there has been 
recent work on partitioning code to find microservices \cite{tysz2018}.  

Most existing work on matching programs concentrates on finding differences rather than 
similarities.  Yang proposed identifying the syntactic differences between two programs by 
matching their syntax trees using dynamic programming \cite{Yang91}.  Neamtiu matches two successive 
programs by visiting their abstract syntax trees in parallel and create maps of their names 
and types \cite{Neamtiu05}.  The tools JDiff \cite{Apiwattanapong04} and UMLDiff \cite{Xing05} focus 
on identifying changes and 
correspondences between object-oriented programs.  LSDiff \cite{Kim09a} identifies program differences 
from the changes of structural dependencies of code elements.  Dex \cite{Raghavan04} creates abstract 
semantic graphs for two versions of programs and then uses the graph differencing algorithm 
to obtain their differences.  Their differencing algorithm iteratively matches graph nodes 
and computes the corresponding costs.  iDiff \cite{Nguyen11} looks at the interaction to compare program 
entities of classes and methods. More recent work looks at finding similar 
software projects \cite{bogomolov2020sosed}. 

The adaptation phases of ASCUS are most closely related to the various refactorings that have 
been proposed and incorporated into various development environments starting with Elbereth \cite{Korman98}.  
Follow-up work looked into how these were used and the problems developers had with 
them \cite{Murphy-Hill08,Murphy-Hill09}.  
This work is also related to the transformations done by the semantic code search tool S6 \cite{scsbcs}. 

There has been significant work on automatically generating test cases.   
white-box techniques look to find method or program inputs that achieve a desired level of coverage.  
Many systems have been proposed and are used in practice, for example, EvoSuite \cite{Fraser11}.    
These systems use a variety of techniques include symbol execution engines and genetic programming.  
Black-box techniques are less common since they typically require a model of the code and generate 
the test cases from the model.  Models have been  based on JML \cite{Cheon07} and 
UML activity diagrams \cite{Samuel09}.  
Black box test cases have also been generated for special cases, for example by monitoring 
program execution \cite{Mariani11}.  Code search has also been used to 
find method-level tests \cite{Reiss14c}. 

\section{Overview}
\begin{figure}[t]
\centering 
\includegraphics[width=\linewidth]{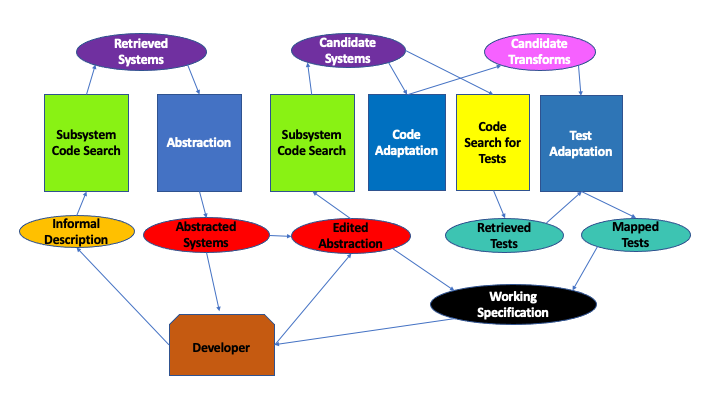}
\caption{Overview of ASCUS.}
\label{fig1}
\end{figure}

Ideally, any subsystem the developer wanted to incorporate would be available as a separate library that could easily
be reused.  In practice, however, these are typically tightly integrated into existing systems require substantial
work to extract, understand, adapt, and reuse.  ASCUS attempts to automate much of this process.  
An overview of the ASCUS approach
can be seen in Figure~\ref{fig1}.

The developer starts by providing an informal description of the subsystem they are interested in.  
This description is used as the basis for a code search from a code repository.   Code search typically will
yield only an individual file, but ASCUS does multiple searches to extract a whole subsystem, find the needed
libraries, and attempt to make the result compilable.  

The returned systems are too complex to be easily understood by a developer.  ASCUS next simplifies them
by abstracting out the essential portions to create a simple, clear interface describing the subsystem.  It
does this using a variety of heuristics based on search terms, visibility, and usage.  These are then presented to the 
developer either as a Java interface or as a UML class diagram.  

The developer can then edit these abstractions to better meet their needs, adding or 
removing classes and methods,
using local types, changing names, etc.  The edited abstractions are then used as the basis for a second
code search for subsystems to yield a set of candidate systems.  The returned subsystems are matched 
against the developer's edited abstraction and a set of transformations
are developed to map the returned code to the abstract specifications.  

Next the project containing each retrieved subsystem is searched for test cases.  These are restricted
to tests of the subsystem and then transformed using the mappings that were computed for the returned code.
The result is a suite of tests for the developer's abstraction.  These are then combined with the abstracted
description to produce a checkable specification that meets the developer's needs.

\section{Searching for Subsystems}

ASCUS's first step is to extract potential subsystems from projects in the repository.  This is done using
the existing repository search facilities which are designed to retrieve projects, files, or methods, not
subsystems.  

Suppose the developer is interested in extending an existing IoT application by adding an embedded http 
server to support a front end that handles static pages for documentation and RESTful requests accessing
data.  They want the footprint of the server to be small. 

ASCUS starts with an informal description from the developer consisting of several sets of keywords.
The first set is optional and is used to identify projects in the repository likely to contain appropriate
code.  We found that searching for projects first and then searching within those projects works better
when searching a large repository such as GitHub.  For the web server example this could be the keywords
\emph{lightweight}, \emph{http} and \emph{server}. 
The second set of keywords is used to search for files.  If no first set was given, then this search 
is over the whole
repository; otherwise it is done once for each of the identified projects.  For the http server
example, this could be just the keyword \emph{server} which would find server instances within the identified
project.

The result of this search is a set of individual files that might be part of an appropriate subsystem.  ASCUS
next expands each of these files into a subsystem by finding other files that are required for compilation
from the same or 
related projects.  It starts by adding all files in the current package that are required to compile the 
original file or any file that is added.  Then it identifies other related packages and adds any required file
from those.  It stops when either too many files have been added (the result is not a subsystem), or when
nothing more can be found.  It then transforms the result so that all the 
defined classes are in
a single package.  

Next, ASCUS filters the results to ensure their relevancy.  It does this using the third set of keywords
denoted as key terms.  These are used to identify potentially relevant fields, methods, and classes.  
Occurrences of these are used to score the resultant subsystems based on relevancy.
For the
http server example, the key terms could be \emph{url, uri, application, property, port, https, ftp, routing,
callback, request} and \emph{response}.  ASCUS filtering removes results that are trivial; that are
overly complex; that are primarily test cases; and that do not meet a minimal level of relevancy 
based on the key terms.  

Finally ASCUS looks at the remaining results, finds references to external packages, and uses these 
to identify any required external libraries by searching in the Maven repository.

\section{Creating Abstractions}

\begin{figure}[t]
  \caption{Sample Java-interface based abstraction.}
  \label{fig2}
  \scriptsize
\begin{lstlisting}
@Ascus(source="GITREPO:https://github.com/Ruhrpottattacke/easyhttpserver/.../HttpServer.java")
@Ascus(library="com.sun.net.httpserver:http:20070405")
@Ascus(library="org.junit.platform:junit-platform-console-standalone:1.8.1")
@Ascus(search="PACKAGE,PACKAGE_USED,500,GITREPO")
@Ascus(keywords={"@lightweight","@http","server"})
@Ascus(keyterms={"server","request","ftp","@lightweight","uri","url","routing",...})
@Ascus(suggestedTerms={"exchange","logger","response","finest","server","handler"})
package edu.brown.cs.SAMPLE;
@AscusPackage
public interface HttpServer {
@AscusClass
interface HttpHandler {
   void handle(HttpExchange exchange);
}
@AscusClass(uses={HttpHandler.class})
abstract class RawHandler implements HttpHandler {
   public RawHandler(HttpHandler handler) { }
   public abstract void handle(HttpExchange exchange);
}
@AscusClass
abstract class HttpServer {
   public HttpServer(int port,int threads,int backlog) { }
   public abstract void start();
   public abstract void stop(int delay);
   public abstract void stop();
   public abstract void createContext(String path,HttpHandler handler);
   public abstract void removeContext(String path);
   public abstract boolean isRunning();
}
@AscusClass
abstract class HttpExchange {
   private String response;
   public HttpExchange(HttpExchange exchange) { }
   public abstract void setResponse(String response);
}
}
\end{lstlisting}
\end{figure}

The subsystems returned by search are still too complex to be quickly understood by developers.  ASCUS
simplifies them by creating abstractions that can be shared with the developer and used as a basis for a
second code search.  The abstractions present the essentials of the subsystem without unnecessary details.
An example of an abstraction can be seen in Figure~\ref{fig2}.

Abstractions are created hierarchically.  First data types are matched for compatibility.  Next fields are
considered in the same field abstraction if their data types are compatible.  Methods are considered in the same
method abstraction if their parameters and return types are compatible disregarding parameter order.  

To abstract a class, ASCUS first identified the relevant fields and methods by 
considering their visibility and the occurrence of key terms in their names or bodies.  Private fields with
a public getter and setter are considered relevant.  The resultant initial abstraction of a class is 
just the set of visible elements.  Similar class abstractions are 
merged by finding the maximal matching of their fields and methods taking into account their abstractions and names.  

Next subsystems are abstracted based on their class abstractions.  The set of relevant classes for a subsystem
starts with all non-trivial, non-private, non-test classes.  Then any subclasses, including classes that implement an 
interface in the set are removed.  Finally, any classes that are referenced by fields or methods in the
abstract class are added back in.  Subsystem abstractions are then merged using a maximal matching with 
a reasonable threshold.

Finally ASCUS adds additional information to the subsystem abstractions that might be helpful to the
developer.  This includes any libraries that are needed to get it to compile and possible additional
keywords for future search based on a tf-idf analysis of the matched code.  Then the
resultant abstractions are then sorted by relevance and presented to the developer.  ASCUS creates
both a single Java interface containing the various elements that can be seen by the developer and a
UML diagram that can be viewed using UMLet or Umbrello. 
For the http server example, ASCUS found 110 possible subsystems in GitHub and returned 18 abstractions
to the developer after filtering including the one shown in Figure~\ref{fig2}.

\section{Matching Abstractions to Retrieved Subsystems}

ASCUS generates abstractions for the developer to edit.  The developer can adapt the abstraction
to their naming conventions and to use their types.  They can remove unneeded methods 
and classes and add other
methods and classes they think are useful.  The resultant edited abstraction then 
forms the syntactic basis of
the specifications for their subsystem.  Once this is done, ASCUS uses the information in the edited 
abstraction as the basis to search for subsystems 
in the repository a second time.  

Once this is done, ASCUS attempts to match the retrieved subsystems with the abstraction and to find
a sequence of code transformations that will map the retrieved code to the specifications of the abstraction.
For each retrieved subsystem it first builds an abstraction of that subsystem.  Then it matches that abstraction
to the developer's edited abstraction.  This first does a matching of data types, then computes maximal matchings
of fields and methods combined with matching words in methods names, comments, and content to find the best
mapping from the retrieved code to the abstraction.  

Then ASCUS computes a sequence of code transformations that will
map the retrieved code to code matching the specification.  These handle renaming, type changes,
parameter order changes, moving classes into or out of other classes, adding missing elements, 
optionally removing unused elements, and ensuring the code obeys the developer's naming conventions.

For the http server example
with relatively simple specifications, ASCUS found and transformed 11 candidate subsystems
ranging in size from 800 to 7000 lines.

\section{Generating Test Cases}

ASCUS next uses the retrieved subsystems that matched the abstraction to find test cases.  It finds tests
for each retrieved subsystem and then combines the resultant tests into a single test suite for the 
abstraction that can be edited by the developer.

To find tests for a retrieved subsystem, ASCUS first does a code search of the repository to find
all files containing JUnit test cases and either an import or package statement from the original retrieved 
code.  It takes all the resultant files and converts them into a set of files in the package associated
with the abstraction.  

It then transforms this file into a set of useful tests.  It starts by applying the transformations
computed by the abstraction matching process to convert types, names, parameter order, etc. from the
original code to the abstraction.  Then it prunes the result by removing any unneeded methods, 
removing any test methods that invoke methods not in the abstraction, and ensuring the remaining
tests compile against the abstraction.  Finally, it transforms the 
result to match the developer's naming conventions.  

The final step in ASCUS's test generation is to combine the results from the different retrieved
subsystems, discard any duplicate tests, and then incorporate the resultant tests into the 
completed abstraction. 

While this step often fails since only about 1/3 of the projects that are retrieved include formal test cases,
it can find actual tests.  For the http server example, about half the subsystems included tests
and ASCUS was able to generate 80 test cases (50 from one subsystem).

\section{Future Plans}

ASCUS currently exists as a proof-of-concept prototype, with no user interface, simple heuristics,
simple matching algorithms, and a small set of code transformations.  Our experiences to date show that 
it is possible to interactively create non-trivial specifications from an informal description in 
under 5 minutes.
However, much needs to be done before this becomes a usable system.

Our short term research plan includes extending the prototype with a suitable web-based user interface;
creating evaluation criteria for the returned abstractions to avoid presenting too many irrelevant
ones to the developers and improve the presentation order; 
improving code search to make it less sensitive to the selection of keywords; and
extending the transformations, heuristics and algorithms used.  
We are also looking at better ways of creating checkable semantics since test cases are not as
common or comprehensive as one would hope.  We are looking at creating test cases from examples of 
how the subsystem
is used in the retrieved package.  We are also looking at other simple means
for letting the developer describe the expected behavior including
UML sequence diagrams and contracts.
Finally, we are investigating different means of evaluating an interactive tool such as
ASCUS.

Checkable specifications of non-trivial subsystems are only a first step.  The ultimate
goal of our research is to generate working versions or at least working skeletons of the described
subsystems.  ASCUS actually does a little of this in transforming the retrieved subsystems to meet
the syntactic specifications of the edited abstraction.  However this is insufficient.

Generating working code that meets a checkable specification will require not only adapting a
retrieved subsystem to the abstraction, but also simplifying that code to remove unnecessary
features; merging multiple retrieved subsystems to provide additional features; editing the
retrieved code to pass the tests, possibly using automatic program repair techniques; generating
code for missing methods using technologies such as program synthesis, code search, or machine
learning; adapting code to use versions of libraries consistent with what the developer is using;
and generally ensuring the code is something the developer actually would want to use.

\clearpage

\bibliographystyle{plain}
\bibliography{paper}

\end{document}